\documentclass[twocolumn]{aastex631}

\usepackage{amsmath}
\usepackage{breqn}

\shorttitle{EM precursors to BH--NS gravitational wave events}
\shortauthors{E.R. Most \& A.A. Philippov}
\graphicspath{{./}{figures/}}

\begin{document}

\title{Electromagnetic precursors to black hole -- neutron star gravitational wave events:\\ Flares and reconnection-powered fast-radio transients from the late inspiral}

\author[0000-0002-0491-1210]{Elias R. Most}
\correspondingauthor{Elias R. Most}
\email{emost@caltech.edu}
\affiliation{TAPIR, MC 350-17, California Institute of Technology, Pasadena, CA 91125, USA}
\author[0000-0001-7801-0362]{Alexander A. Philippov}
\affiliation{Department of Physics, University of Maryland, College Park, MD 20742, USA}

\begin{abstract}
  The presence of magnetic fields in the late inspiral of black hole -- neutron star binaries could lead to potentially detectable
  electromagnetic precursor transients. Using general-relativistic force-free electrodynamics simulations, we investigate pre-merger interactions of the common magnetosphere of black hole -- neutron star systems.  
  We demonstrate that these systems can feature copious electromagnetic flaring activity, which we find depends on the magnetic field orientation but not on black hole spin. Due to interactions with the surrounding magnetosphere, these flares could lead to Fast Radio Burst-like transients and X-ray emission, with $\mathcal{L}_{\rm EM} \lesssim 10^{41} \left( B_\ast/ 10^{12}\, \rm G \right)^2\, \rm erg/ s$ as an upper bound for the luminosity, where $B_\ast$ is the magnetic field strength on the surface of the neutron star.
\end{abstract}

\keywords{Black holes(162), General relativity(641), Gravitational wave sources (677), Neutron stars (1108), Radio
transient sources (2008), Radio bursts(1339), Magnetospheric radio emissions (998), Plasma
astrophysics(1261), X-ray transient sources(1852)}

\section{Introduction}

The first detections of gravitational waves (GW) from two black hole
(BH) -- neutron star (NS) mergers by the LIGO-VIRGO-KAGRA collaboration
\citep{LIGOScientific:2021qlt} have provided us with a first glance at the merging NS-BH binary population. 
So far these systems have featured
BH masses in excess of $5\, M_\odot$ with moderate dimensionless spins, $\left|\chi \right|\lesssim 0.3$ \citep{LIGOScientific:2021qlt}.
Such systems have been shown to only leave negligible amounts of remnant
baryon matter after the merger, if any \citep{Foucart:2018rjc,Foucart:2012nc}, potentially making them less likely sources for multimessenger electromagnetic (EM) counterparts \citep{Fragione:2021cvv,Biscoveanu:2022iue}, such as those observed for the NS-NS event GW170817 (e.g., 
\citealt{Cowperthwaite:2017dyu,Chornock:2017sdf,Villar:2017wcc,Nicholl:2017ahq,Troja:2018ruz,Tanvir:2017pws,Drout:2017ijr,LIGOScientific:2017zic,Savchenko:2017ffs,Troja:2017nqp,Margutti:2017cjl,Margutti:2018xqd,Hajela:2019mjy,Hallinan:2017woc,Alexander:2017aly,Ghirlanda:2018uyx,Mooley:2017enz,Mooley:2018qfh}).
Indeed, none have been observed for any of the current BH-NS events
\citep{Anand:2020eyg,Coughlin:2019zqi,Coughlin:2020fwx}, see also
\citet{Raaijmakers:2021slr}. At the same time, observing faint counterparts might
crucially rely on the ability to perform rapid sky localizations, which could be aided by all-sky radio observations \citep{Sachdev:2020lfd,Yu:2021vvm}. This makes understanding the possibility of other, not yet observed, transients that could potentially be sourced before or right during the merger even more important.
In fact, prior to the merger, the orbiting neutron star can feature a strong exterior magnetic field, whose dynamics could be relevant in sourcing additional EM transients
\citep{Hansen:2000am,Lyutikov:2018nti}.  
Indeed, this possibility was investigated for previous gravitational wave events
\citep{Callister:2019biq,Broderick:2020lcv}, see also
\citep{Stachie:2021noh,Stachie:2021uky}, with further efforts being proposed
for future searches
\citep{James:2019xca,Sachdev:2020lfd,Yu:2021vvm,Wang:2020sda,Gourdji:2020rca,Cooper:2022slk}.
In the context of BH-NS GW events \citep{LIGOScientific:2021qlt}, the nondetection of such precursor counterparts
was used to constrain the magnetic field strength present in the stars before merger \citep{DOrazio:2021puy}.

\begin{figure}
    \centering
    \includegraphics[width=0.45\textwidth,trim={15cm 10cm 25cm 10cm},clip]{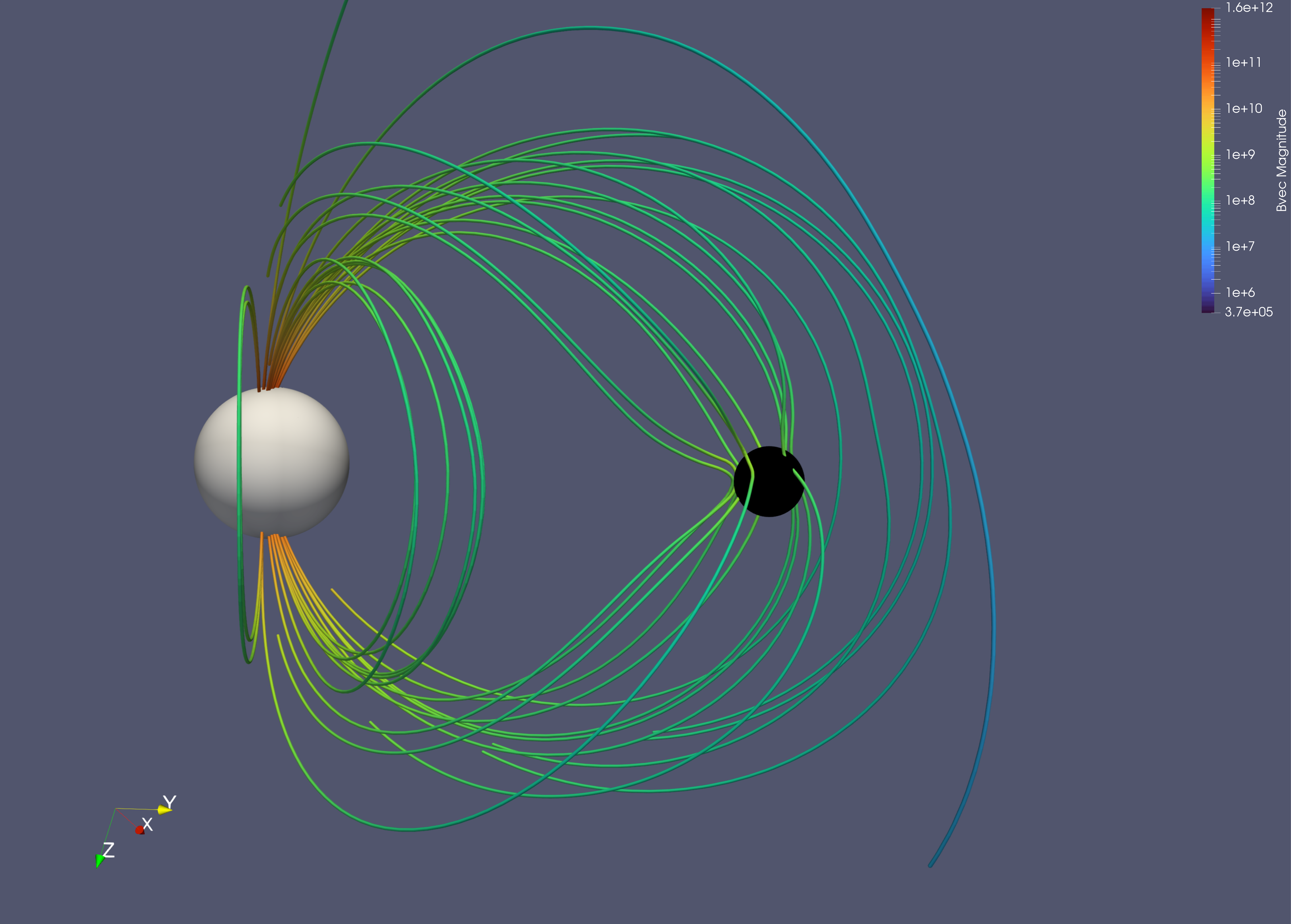}
    \caption{Interaction of a black hole {\it (right)} with the magnetic field lines of a neutron star {\it (left)} shortly before merger.}
    \label{fig:intro}
\end{figure}

In order to predict what type of precursor transient to expect, it is 
necessary to clarify the various production mechanisms potentially operating in the binary magnetosphere. 
One class of scenarios is concerned with transients produced during merger,
when the magnetosphere of the NS transitions onto the BH
\citep{DOrazio:2013ngp,Mingarelli:2015bpo,DOrazio:2015jcb},
see also \citet{Nathanail:2020fkp} for cases involving NS-NS prompt collapse. 
The newly created magnetic field topology on the BH is not stable and will
cause a dynamical transient associated with both BH ringdown
\citep{Lehner:2011aa,Palenzuela:2012my,Dionysopoulou:2012zv,Nathanail:2017wly,Most:2018abt},
and later magnetic balding of the BH \citep{Bransgrove:2021heo}, see also \citealt{East:2021spd} for a recent simulation in dynamical spacetime, which can result in formation of strong shock waves and a bright X-ray signal \citep{Beloborodov:2022pvn,Beloborodov:2023lxl}.\\
Apart from these violent one-off transients, orbital motion and interactions inside the common magnetosphere can also drive
periodic EM outflows. In particular, \citet{McWilliams:2011zi,Lai:2012qe,Piro:2012rq} have considered energy extraction and predictions for EM precursors due to the interaction of the fields with an orbiting  companion, while \citet{Sridhar:2020uez} have considered shock-powered precursors from winds. Numerically, some of these cases have been investigated by
\citep{Paschalidis:2013jsa,Carrasco:2019aas,Carrasco:2021jja}, see also
\citet{Palenzuela:2013kra,Palenzuela:2013hu,
Ponce:2014hha} for the case of
NS-NS mergers. Other works have also considered the impact of net stellar or black hole charge potentially building up during the inspiral, which in part may depend on NS spin \citep{Levin:2018mzg,Zhang:2019dpy,Dai:2019pgx}.
In the scenario we will consider in this work, we focus on interactions in
the common magnetosphere that will naturally and generically lead to periodic emission of powerful EM
flares prior to merger \citep{Most:2020ami}, akin to coronal mass ejections
in the Sun \citep{2011LRSP....8....1C}. In the case of coalescing NS-NS
binaries, these happen for a large class of magnetic field geometries and
field strength values, with about 10-20 observationally significant flares being
launched prior to merger (\citealt{Most:2022ojl}, assuming magnetic field
strength of at least $10^{11}\, \rm G$ at the surfaces of the stars).
Observationally, these flares have been predicted to give rise to
Fast-Radio-Burst-like transients at higher frequencies ($10-20\, \rm GHz$) with luminosities  $\mathcal{L} \lesssim 10^{42}\, \rm erg/s$.
\citep{Most:2022ayk}.\\ 
For BH--NS systems, we demonstrate that precursor
flares will always be emitted as long as the magnetic field inclination of the
NS exceeds a critical limit. 
More specifically, we investigate a variety of magnetic
field topologies, orbital separations and BH spins (see Fig.
\ref{fig:intro}). We describe our computational and numerical approach in
Sec. \ref{sec:methods}. In Sec. \ref{sec:results}, we then demonstrate that
BH-NS systems can periodically emit powerful flares.  In Sec.
\ref{sec:budget}, we quantify the energy budget of these flares and discuss
implications for potential fast radio and X-ray transients.

~\\~\\

\section{Methods} \label{sec:methods}
In this work, we model the magnetospheric dynamics of a BH--NS system. 
To do so, we need to take two aspects of the system into account:
The plasma inside the magnetosphere, and the relativistic effects of the
system, in particular, the BH and its ergosphere.\\
Sourced by the magnetic field of the NS, the orbital motion will
trigger pair creation prior to merger
\citep{Hansen:2000am,Lyutikov:2018nti} making the magnetosphere nearly force-free
\citep{Goldreich:1969sb}. We can model its dynamics by adopting the force-free electrodynamics approximation
\citep{Palenzuela:2010nf,Parfrey:2013gza,Carrasco:2019aas}. Such 
an approximation is well justified in a closely orbiting system with
sufficiently large magnetic fields
\citep{Hansen:2000am,Lyutikov:2018nti,Wada:2020kha},
and has been used extensively to study the pre-merger magnetosphere of
compact binaries \citep{Palenzuela:2012my,Dionysopoulou:2012zv,Paschalidis:2013jsa,Palenzuela:2013hu,Palenzuela:2013kra,Ponce:2014hha,East:2021spd,Carrasco:2021jja}. We specifically make use of the formulation
of \citet{Palenzuela:2013hu} as used in our previous studies \citep{Most:2020ami,Most:2022ojl,Most:2022ayk}.

The force-free electrodynamics conditions \citep{Gruzinov:1999aza} are enforced using an
effective parallel conductivity term \citep{Alic:2012df,Spitkovsky:2006np}.
We then solve the discrete form of the general-relativistic force-free electrodynamics system
using a high-order finite-volume scheme
\citep{mccorquodale2011high} with WENO-Z reconstruction \citep{Borges2008}. 
The numerical code is implemented on top of the
adaptive mesh-refinement infrastructure of the \texttt{AMReX} framework
\citep{amrex}. More details on this implementation can be found in
\citet{Most:2022ojl}.

Secondly, we need to model the orbiting binary system, including the near-horizon geometry of the BH. 
Several approaches have been proposed
in the literature. It is possible to model the BH as a resistive
surface using the approximate membrane-paradigm \citep{Thorne:1986iy}. Such approaches have
been used in flat spacetime to model BH flares from interactions with coronal
magnetic fields \citep[e.g.,][]{Yuan:2019mdb}. At the same time, these can only
approximately capture the ergospheric dynamics operative in the BH
magnetosphere \citep{Komissarov:2004ms}. Alternatively, it has been successfully demonstrated
that the common magnetosphere of the system can be modeled as a Kerr
spacetime \citep{Kerr:1963ud} with the NS (modeled as a spherical conductor) moving on a point-particle orbit
\citep{Carrasco:2021jja}. While this captures the necessary BH dynamics, it only
approximately models the orbit and the NS. Finally, it would also
be possible to resort to fully dynamical spacetime simulations of the
system \citep{East:2021spd}. While these can capture all the necessary aspects mentioned
above, solving the Einstein equations is computationally expensive. It typically prevents the use of very high numerical resolution necessary to
avoid spurious diffusion in the magnetosphere. \\
In this study, we use a third way, in between a stationary
spacetime approach and a full numerical relativity simulation.
Following \citet{Paschalidis:2013jsa}, we make use of numerical relativity initial conditions
for such a system, constructed by imposing a helical Killing vector, i.e., constant orbital separation \citep{Grandclement:2006ht,Taniguchi:2007xm,Taniguchi:2007aq,Foucart:2008qt}.
\begin{table}
  \centering
\begin{tabular}{|l|c|c|c|c|}
    \hline
    \hline
    &$\Omega$ [$\rm s^{-1}$]  & $a\, \left[\rm km\right]$ &
    $\chi_{\rm BH}$ & $\theta_{\rm NS} [^\circ]$\\
    \hline
\hline
        NSBH\_$\chi$085\_60\_60 & 925.3.0& 103.3& 0.85 &  60\\
	NSBH\_$\chi$085\_60\_70 & 765.0& 103.3& 0.85 &  60\\
 	NSBH\_$\chi$085\_60\_80 & 637.0& 118.1& 0.85 &  60\\
\hline
   	NSBH\_$\chi$-06\_60 & 928.2 & 89.9 & -0.60 &  60\\
 	NSBH\_$\chi$-03\_60 & 928.2 & 89.9 & -0.30 &  60\\
 	NSBH\_$\chi$0\_60 & 928.2 & 89.9 & 0.00    & 60\\
   	NSBH\_$\chi$03\_60 & 928.2 & 89.9 & 0.30   & 60\\
 	NSBH\_$\chi$85\_60 & 928.2 & 89.9 & 0.85   & 60\\

\hline
        GW200115$^\ast\_0$ & 996.7 &89.9 & -0.30  &   0\\
        GW200115$^\ast\_30$ & 996.7 &89.9 & -0.30 & 30\\
        GW200115$^\ast\_45$ & 996.7 &89.9 & -0.30 & 45\\
        GW200115$^\ast\_60$ & 996.7 &89.9 & -0.30 & 60\\
        GW200115$^\ast\_90$ & 996.7 &89.9 & -0.30 & 90\\
\hline
\hline
  \end{tabular}
	\caption{Summary of the black hole (BH) -- neutron star (NS) binaries considered in this work. The columns denote 
        the orbital angular frequency $\Omega$,
	  the orbital separation $a$, the dimensionless BH spin $\chi_{\rm BH}$ perpendicular to the orbital plane, and the NS dipole inclination relative to the orbital angular momentum axis. 
        We further adopt canonical masses of $m_{\rm BH} = 5 \, M_\odot$ and $m_{\rm NS} = 1.4\, M_\odot$ for the BH and NS, respectively. The NS is assumed to be irrotational. Because of the decoupling of electromagnetic and gravitational sectors in our setup, the simulation results are scale-invariant under the magnetic field strength. For our fiducial GW200115-like models we adopt negative spins consistent with the original detection paper \cite{LIGOScientific:2021qlt}, but caution that their astrophysical interpretation remains uncertain \citep{Mandel:2021ewy}. }
    \label{tab:initial}
\end{table}

\begin{figure*}
    \centering
\includegraphics[width=\textwidth]{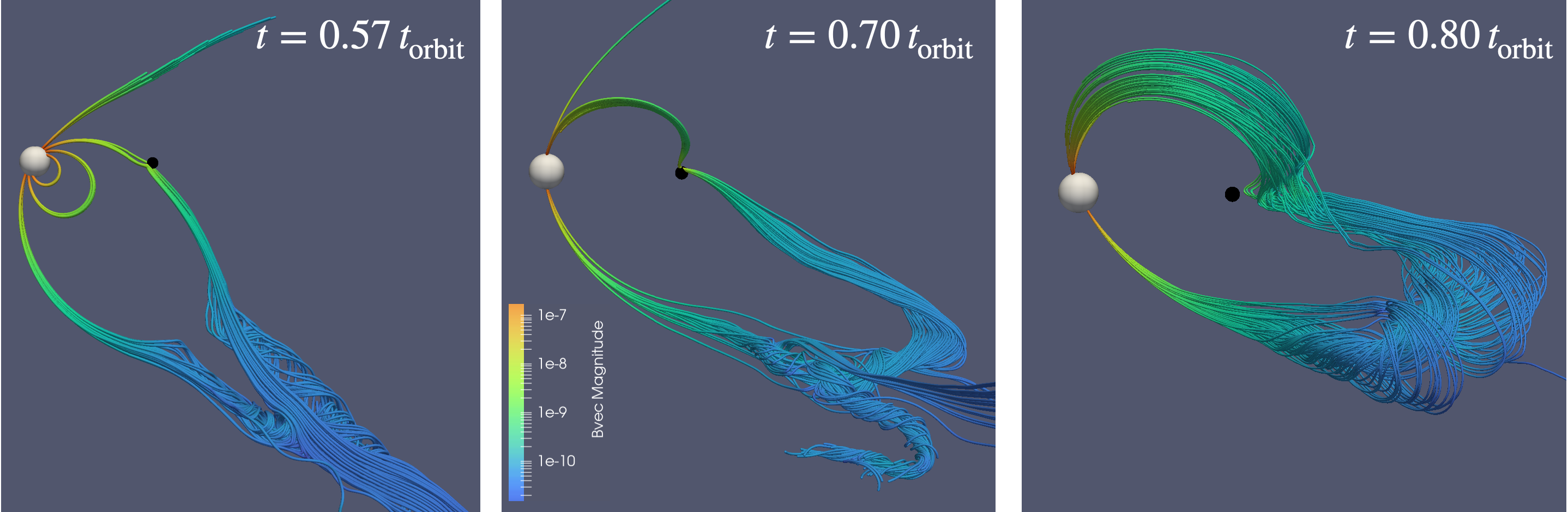}
    \caption{Three-dimensional magnetic field line visualization of a
      flaring event for a GW200115-like binary with a magnetic dipole with
      $\theta_{\rm NS} = 60^\circ$ inclination. Shown are selected field
      line bundles for three different times, $t$, relative to the orbital
      period, $t_{\rm orbit}$. {(\it Left)} Closed magnetic field lines get
      initially accreted and pinned onto the black hole. {(\it Center)} Continued orbital motion begins twisting the connected field lines, inflating the connected flux tube.
    {(\it Right)} The flare detaches from the binary as can be seen by field lines disconnecting from the black hole (due to reconnection in trailing current sheet). The strong twist is visible as a toroidal component of the inflated flux tube. Colors indicate logarithmic magnetic field strengths in relative units.}
    \label{fig:GW200115_3D}
\end{figure*}
More concretely, we solve the extended-conformal-traceless-sandwich 
formulation for a BH--NS system \citep{Taniguchi:2007aq,Tacik:2016zal} with Cook-Pfeiffer boundary conditions on the black hole \citep{Cook:2004kt,Caudill:2006hw}.
We highlight that this choice of initial data allows us to perform
simulations at fixed orbital separation in full general relativity, representing
the most natural extension of our flat spacetime setup to full
general-relativity \citep{Most:2022ojl}.
Numerically, this system is solved using the \texttt{FUKA} code \citep{Papenfort:2021hod}, which operates on top of the spectral
\texttt{Kadath} framework \citep{Grandclement:2009ju}. The resulting initial data is directly
interpolated onto our computational grid using the spectral representation.
The interior of the black hole is filled using eighth-order Lagrange
extrapolation \citep{Etienne:2007hr}.\\
To evolve these initial conditions using force-free electrodynamics we need to perform two
modifications. First, we need to identify the surface of the
NS in order to impose boundary conditions for the force-free
electrodynamics evolution of the magnetosphere. This is done approximately
by identifying a level surface of the rest-mass density, $\rho\simeq
10^{-6} \rho_{\rm max}$, where $\rho_{\rm max}$ is the maximum rest-mass
density inside the neutron star. \\
Second, following our previous approach in flat spacetimes \citep{Most:2022ojl}, we solve the extended Maxwell system
also inside the neutron star by imposing ideal magnetohydrodynamics
conditions on the electric field, ${\bf E} = - {\bf v} \times {\bf B}$.
This requires a prescription for the three-velocity, ${\bf v}$, of the
spacetime foliation. In principle, the velocity field can be self-consistently
obtained from the numerical initial conditions. However, it is currently not known how to self-consistently impose
a velocity field corresponding to perfect uniform rotation \citep{Tichy:2011gw,Tichy:2012rp}.
Indeed, we have found that when evolving these consistent velocity profiles
small non-uniformities in the electric field close to the surface lead to a spurious diffusion of the magnetic field out of the NS. This motivated us to manually replace the velocity profile
with that of perfect uniform rotation consistent with the rotation frequency of star
prescribed in the initial conditions. Since we are keeping both spacetime and hydrodynamics fixed, 
this amounts to merely modifying the boundary conditions at the surface of the star. Indeed,
we could equivalently have opted to excise the stellar interior and impose those
boundary conditions only on the surface instead \citep{Carrasco:2019aas}.

\section{Results}
\label{sec:results}

In the following, we will present our results on the evolution of the joint BH-NS magnetosphere and electromagnetic precursor
flares emitted from them. These are based on performing systematic
numerical studies of general-relativistic BH-NS magnetosphere dynamics for
the models listed in Tab. \ref{tab:initial}. While we will mainly focus on a 
fiducial system consistent with the GW200115 event \citep{LIGOScientific:2021qlt},
we will also present a parameter study in terms of BH spin and binary separation, allowing us
to generalize our findings to arbitrary systems with realistic mass ratios.\\
The present work is divided into several parts.  
First, we begin by describing the flares and flaring mechanism operating inside the BH--NS magnetosphere (Sec. \ref{sec:GW200115}). We then discuss its dependence on the magnetic field topology (Sec. \ref{sec:inclination}), and finish with a discussion on potentially observable signatures of these flares (Sec. \ref{sec:budget}).

\subsection{Precursor flares from the late of inspiral of black hole --
neutron star binaries}\label{sec:GW200115}

In this Section, we will describe the generic flaring mechanism operating
in the magnetosphere of a GW200115-like system.\\
We start out with a description of the common pre-merger magnetosphere.
In this regime, the magnetic field lines of the
NS can thread the BH, establishing a connected flux tube between the two
compact objects in the binary system (see Fig. \ref{fig:intro}). This process
will happen regardless of the orientation of the NS magnetic field. This is
in contrast to NS-NS pre-merger magnetospheres
\citep{Most:2020ami,Most:2022ojl}, where a minimal degree of misalignment
between the two stellar magnetic moments is required in order to establish connected
flux tubes \citep{Cherkis:2021vto}. Orbital motion of the system can then
lead to a build-up of twist in these connected loops, causing them to
inflate \citep{Most:2020ami}. However, unlike for the NS, which is a perfect
conductor, the BH allows for field line slippage
\citep{MacDonald:1982zz,Thorne:1986iy}. Furthermore, since a BH cannot
support closed field lines anchored solely in it \citep{MacDonald:1982zz},
the effective field geometry on the BH will have both open (split-monopole-like) and closed (flux tubes connected to the NS) field lines. Within this
magnetosphere, a current sheet will form, causing electromagnetic energy to be dissipated. The extent of this current sheet is larger than that of a rotating BH \citep{Komissarov:2007rc}. We attribute this to the presence of the orbital motion making the local magnetosphere of the BH more closely resemble that of a boosted BH configuration \citep{Carrasco:2019aas}. 
The energetics of the dissipation in the individual magnetospheres have been extensively studied by \citet{Carrasco:2021jja}, and we will not
focus on those aspects associated with the energy dissipation of each
individual compact object here. Rather, we will describe the dynamics of
the common flux tube, which -- as we will demonstrate -- is able to give rise to realistic scenarios for EM transients, akin to NS-NS magnetospheres \citep{Most:2022ayk}.\\
The general dynamics of the connected flux tube in the BH-NS pre-merger
magnetosphere is shown in Fig. \ref{fig:GW200115_3D}. There, we focus on the  dynamics of a GW200115-like binary with a NS magnetic moment
inclined by $\theta_{\rm NS} = 60\,^\circ$ relative to the orbital angular momentum axis (see Tab. \ref{tab:initial} for details).  While flaring will happen periodically, with about two
flares being emitted per orbit \citep{Cherkis:2021vto}, we will first focus
on a single flaring event.\\
Starting with the situation shown in the left
panel of Fig. \ref{fig:GW200115_3D}, we can see that due to orbital motion
the closed zone of the NS magnetosphere will start to interact the BH. As we show in Sec. \ref{sec:inclination}, it will be crucial
that the BH interacts with field lines forming the outer part of the
magnetosphere, i.e., field lines being close to the orbital light cylinder. As the
binary continues to orbit, an effective twist is built-up
and the connected flux tubes begin to inflate (Fig. \ref{fig:GW200115_3D},
middle panel). We can see that the dynamics of connected field lines closer to the orbital light cylinder differ from those further inwards . This is mainly a result
of the different field strengths of the flux tubes connecting the NS and BH. 
While the shorter flux tube slowly inflates, the larger flux tube
extends further outwards. As indicated by the color coding, the smaller
flux tube (green color) has higher field strengths compared to the lower
one (blue color).  This means that the same twist applied to each flux tube
will lead to a different morphology. As can be seen in the final panel of
Fig. \ref{fig:GW200115_3D}, the longer flux tube gets strongly twisted and
contains a strong toroidal field component, visually showing the twist. It
is this outer part at lower field strengths (blue color), which will detach
and flare. The upper loop does not show any deformation. Instead, because of dissipation in the
trailing current sheet \citep{Lyutikov:2011tk}, we observe a steady state
twist in this case. Since force-free models do not support plasma pressure, current sheets may collapse quicker and dissipate more, compared to current sheets in a fully kinetic description\footnote{Force-free current sheets, such as the ones studied here, tend to be very resistive with field lines closing through the current sheet. This is in contrast to magnetohydrodynamic \citep{Komissarov:2005wj} or particle-in-cell studies \citep{Parfrey:2018dnc} of reconnecting black hole magnetospheres. }.\\
This lets us draw an important conclusion.  While BH-NS systems can
naturally flare, the flaring will likely release energy build-up around the orbital light cylinder (i.e., in the lower field
strengths part of the magnetosphere). This is different from NS-NS systems,
in which flux tubes deep insight the orbital light cylinder will flare \citep{Most:2022ojl}. 
The amount of energy released in a flare (or even whether
flaring happens at all), will then depend on the relative inclination of
the magnetic moment of the star relative to the orbital rotation axis. We
will quantify this in Sec. \ref{sec:inclination}.

\begin{figure}
    \centering
    \includegraphics[width=0.45\textwidth,trim={2cm 2cm 2cm 2cm},clip]{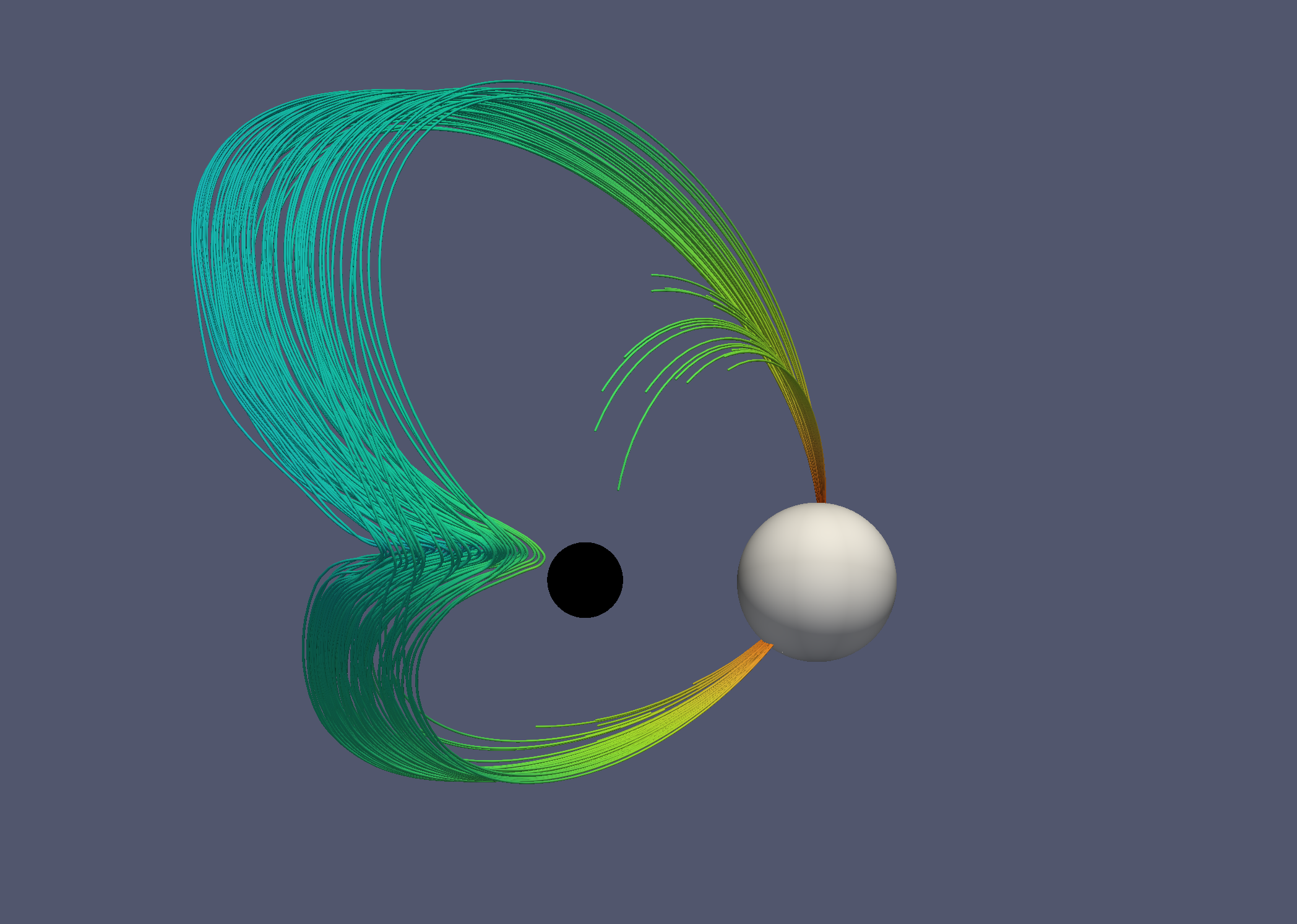}
    \caption{Three-dimensional magnetic field line visualization of a flaring event for a GW200115-like binary for a magnetic dipole with $\theta_{\rm NS} = 30^\circ$ inclination at a time $t \simeq t_{\rm orbit}$, where $t_{\rm orbit}$ is the orbital period. Shown are selected field line bundles that highlight the presence of the equatorial current sheet. Unlike larger magnetic dipole inclinations, efficient reconnection in this current sheetself-limits the twist imposed on the flux tubes. The configuration shown here is in a quasi-steady state, and no flaring is observed.}
    \label{fig:inclination}
\end{figure}

\subsection{Importance of magnetic field inclination}\label{sec:inclination}
\begin{figure*}
    \centering
    \includegraphics[width=0.49\textwidth]{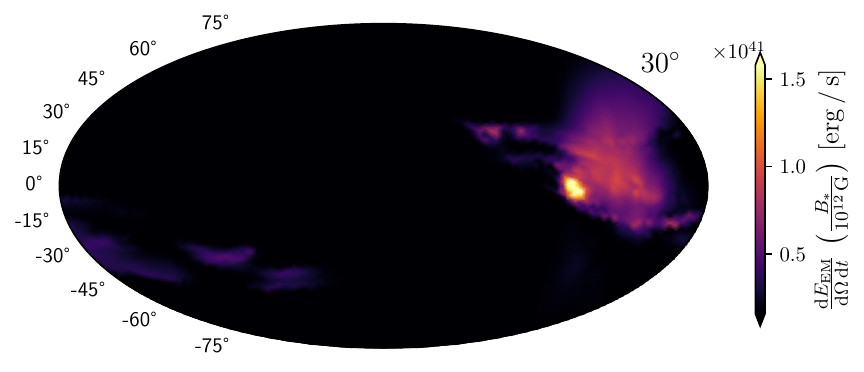}
    \includegraphics[width=0.49\textwidth]{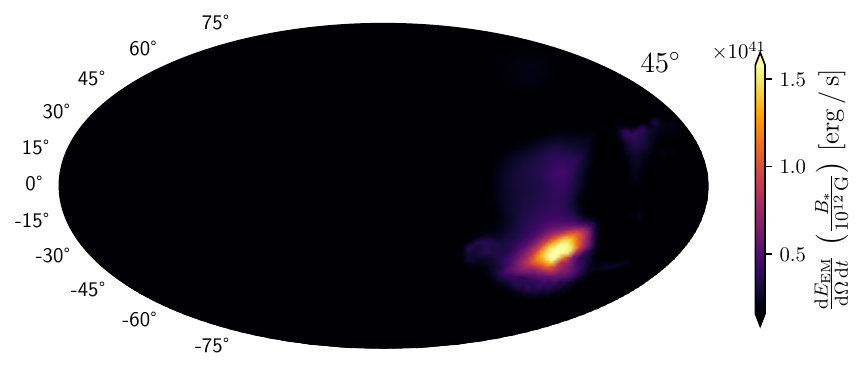}\\
    \includegraphics[width=0.49\textwidth]{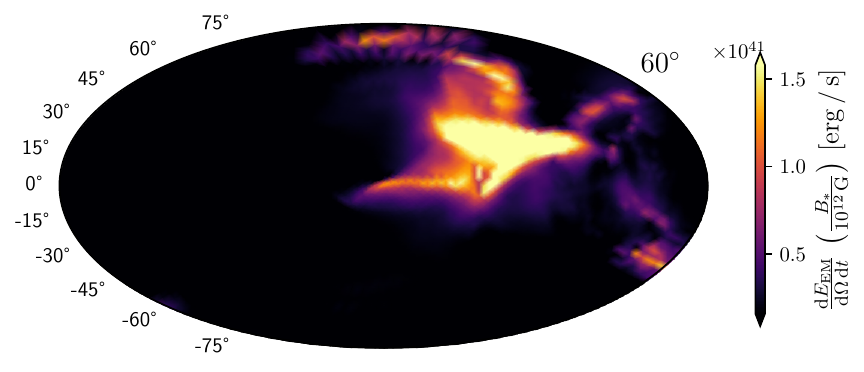}
    \includegraphics[width=0.49\textwidth]{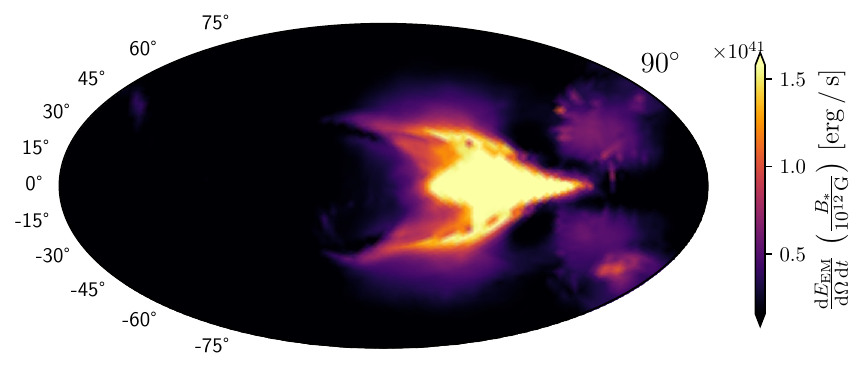}
    \caption{Spherical projection of the average outgoing electromagnetic
      energy flux density of the flares for various inclinations, $\theta_{\rm NS}$, of the neutron
      star's dipole magnetic field extracted at a radius $r=500\, \rm km$. The background flux has been subtracted.
    From top left to bottom right, we show cases for $\theta_{\rm NS} = 30^\circ$ {\it (top
    left)},  $45^\circ$ {\it (top right)},  $60^\circ$ {\it (bottom left)}
  and $90^\circ$ {\it (bottom right)}.}
    \label{fig:2D_poynting}
\end{figure*}
\begin{figure}
    \centering
    \includegraphics[width=0.48\textwidth]{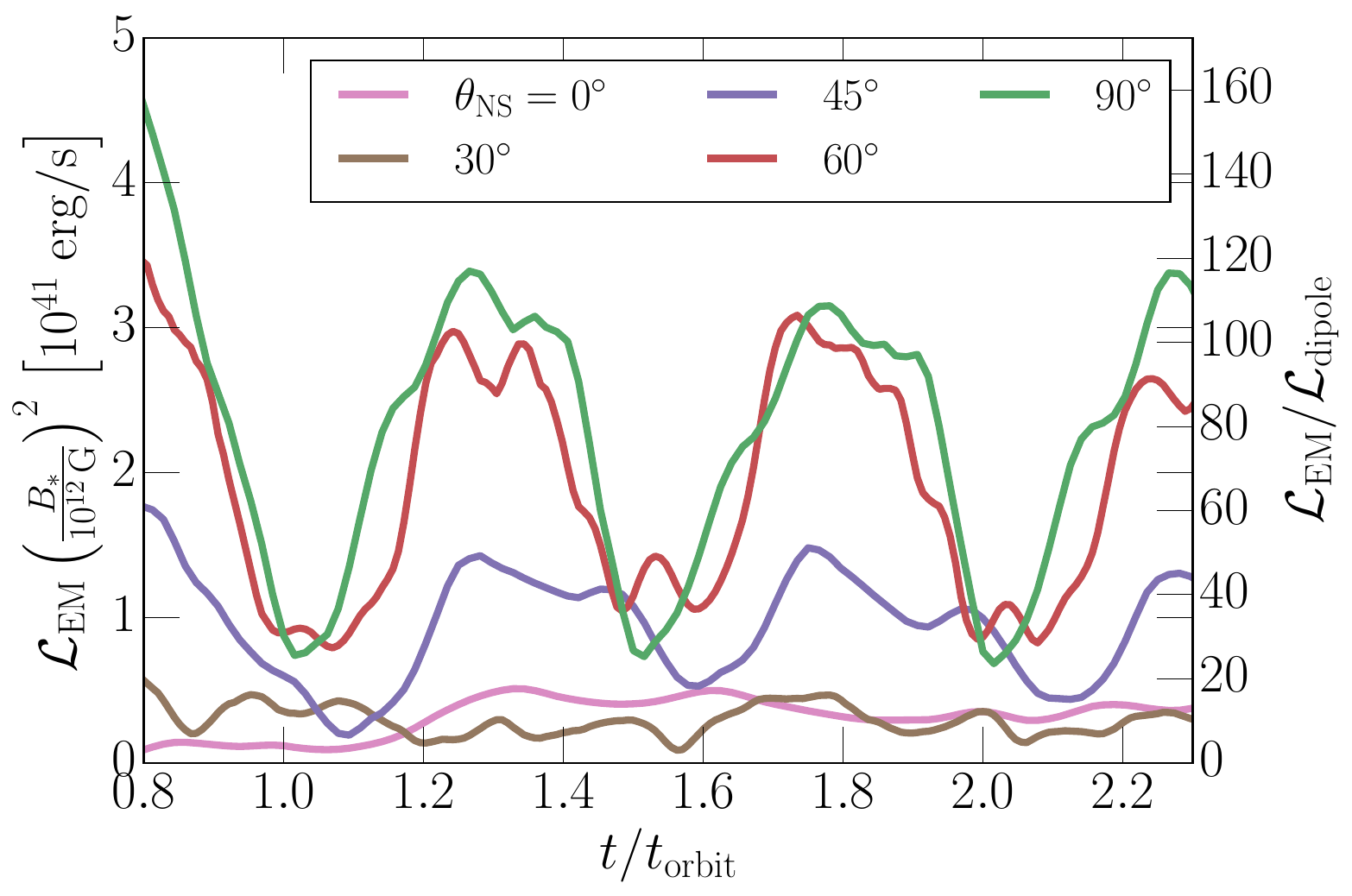}
    \caption{Electromagnetic luminosities, $\mathcal{L}_{\rm EM}$, for GW200115-like models for various inclination angles, $\theta_{\rm NS}$, of the neutron star's magnetic field. Normalizations are given relative to the moving dipole estimate, $\mathcal{L}_{\rm dipole}$, see also \eqref{eqn:orbiting_dipole}, as well as to the orbital period, $t_{\rm orbit}$. All luminosities have been computed at a distance $r=500\,\rm km$ from the system.}
    \label{fig:L_EM}
\end{figure}

Having described the flaring mechanism operating in a BH-NS magnetosphere
prior to the merger, we now proceed with describing the characteristic
properties of the flares. In doing so, we will highlight the importance of the
inclination of the NS magnetic moment for the flaring mechanism.  \\
Since there are no direct a priori constraints on the orientation of premerger magnetic
fields, we consider a range of magnetic field inclinations
$\theta_{\rm NS} = [0^\circ, 30^\circ, 45^\circ, 60^\circ, 90^\circ]$. Statistically, this angle of inclination is expected to correlate with the age of the pulsar \citep{1998MNRAS.298..625T,2010MNRAS.402.1317Y}.
All models are listed in Table \ref{tab:initial}. 
Depending on this inclination angle, we find two general outcomes:
For inclinations above $45^\circ$ our simulations reveal a flaring state as
described before and shown in Fig. \ref{fig:GW200115_3D}.  For inclinations
$\theta_{\rm NS} \lesssim 45^\circ$, we, on the other hand, find no flaring.
Instead, a constant state of dissipation in the current sheet around the BH sets in,
leading to the build-up of a steady twist, as shown in Fig.
\ref{fig:inclination}. We point out that this is likely enhanced by the fact that the current sheet is enlarged due to orbital motion, compared to an isolated spinning BH.
Compared with Fig. \ref{fig:GW200115_3D}, where
flaring occurred at low field strength field lines (blue color), the
$\theta_{\rm NS} = 30^\circ$ system (see Fig. \ref{fig:inclination}) twists
only inner field lines at higher field strengths (green colors). Similar to the
shorter twisted flux tubes in Fig. \ref{fig:GW200115_3D}, the orbital
motion is insufficient to cause these field lines to inflate and twist.
Indeed, this system (and any system at lower inclination) does not flare.
This situation is loosely similar to the polar flaring dynamics of magnetar explosions
\citep{Parfrey:2013gza,Carrasco:2019aas}. For
these, \citet{Mahlmann:2023ipm} have recently shown that depending on the
latitude of the footpoint of the twisted flux tube, it will either flare or
become kink-unstable, thereby dissipating part of the twist. The distinction
between the two states directly results from the different locations inside the magnetosphere that experience the twist.
Polar field lines experience less pressure from confining magnetospheric field lines, and will more easily flare, i.e., produce an electromagnetic outflow.
In the case of the BH-NS systems
considered here, the role of latitude is played by the inclination angle of
the NS's magnetic moment.  In summary, we find that a minimum
inclination angle $\theta_{\rm NS}\simeq 45^\circ$ may be necessary for
flaring, although the precise behavior will also depend on the dissipation rate inside the current sheet. In this sense our results should be interpreted as a lower limit on the amount of flaring. \\
The results of a parameter survey for these systems are shown in Fig.
\ref{fig:2D_poynting}, which show the time-integrated energy fluxes per
spherical angle on a sphere at radius $r\,\simeq 500\, \rm km$ from the
orbital axis. Rather than showing the flux for a given time, we have
averaged the Poynting flux density over one flaring event.
Beginning with the $\theta_{\rm NS}= 30^\circ$ case (top left) we can see that no clear
flaring structure is present. There seems to be a tiny amount of residual
emission associated with steady-state dissipation, potentially accompanied
by the ejection of large plasmoids \citep{Carrasco:2021jja}.
Beginning with $\theta_{\rm NS} = 45^\circ$, we instead see a clear flare
structure emerging, i.e. a localized energy flux at high lattitude. The
precise lattitude of the flare correlates directly with $\theta_{\rm NS}$,
with flares being progressively emitted closer to the equator with
increasing inclination, $\theta_{\rm NS}$. At the same time, we can also
clearly observe an increase in strength and size of the flares for more
strongly misaligned systems.\\
We can furthermore quantify the overall luminosity of the flares by
integrating the previously shown Poynting flux densities over the spheres.
In Fig. \ref{fig:L_EM}, we can indeed see that the system flares twice per
orbit. Moreover, we confirm that the luminosity of more strongly inclined
flares can be enhanced by a factor two, and that no detectable flaring is
present in our simulation at small inclinations.

To better understand the expected behavior we need
to establish a baseline for the luminosity emitted in these systems.
We therefore compare the luminosity of the flares to the one emitted by the
orbital motion of a vacuum dipole.
This has been found to be \citep{Hansen:2000am,Ioka:2000yb},
\begin{align}
    \mathcal{L}_{\rm Dipole} &= \frac{4}{15c^5} \mu^2 a^2 \Omega^6\,,  \label{eqn:orbiting_dipole}\\
     &\approx 2.9\cdot 10^{39} \frac{{\rm erg}}{\rm s} 
    \left(\frac{B_\ast}{10^{12}\, \rm G}\right)^2 \left(\frac{M}{6.4\, M_\odot}\right)^3 
    \left(\frac{90\, \rm km}{a}\right)^7 \nonumber\,,
\end{align}
where $B_\ast$ is the magnetic field strength at the neutron star surface,
$M$ is the total mass of the system, and $a$ the binary separation. In
simplifying the expression we have further assumed that (neglecting
gravitational radiation reaction) the binary is on a Keplerian orbit
\citep{Peters:1964zz}, and that the magnetic moment $\mu = B_\ast
R_\ast^3$, where $R_{\ast} = 13\, \rm km$ (e.g., \citet{Ozel:2016oaf}).
Based on this estimate, the flares present in our simulations are up to a $100$-times
more luminous than the expected orbital emission.
On the contrary, our non-flaring simulations are indeed roughly
consistent with the estimate \eqref{eqn:orbiting_dipole}, in line
with results reported by \citet{Carrasco:2021jja}, who found an additional enhancement due to relativistic orbital motion of the NS. Overall, the
luminosities expected for a surface magnetic field $B_\ast \simeq 10^{12}\,
\rm G$ magnetic field will be roughly comparable to luminosities
$\mathcal{L}_{\rm EM} = 10^{41}\, \rm erg /s$.\\

\begin{figure}
    \centering
    \includegraphics[width=0.47\textwidth]{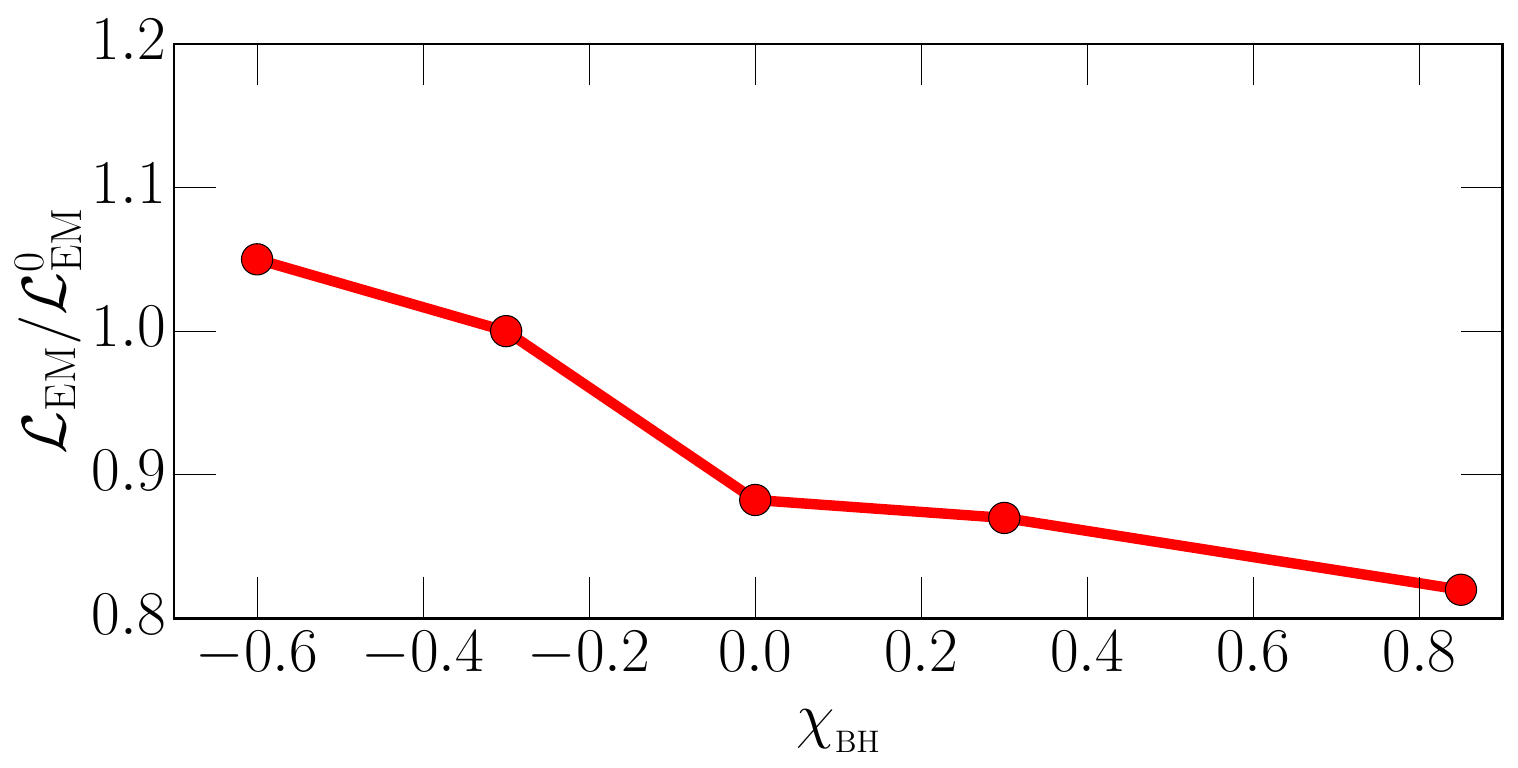}\\
    \includegraphics[width=0.47\textwidth]{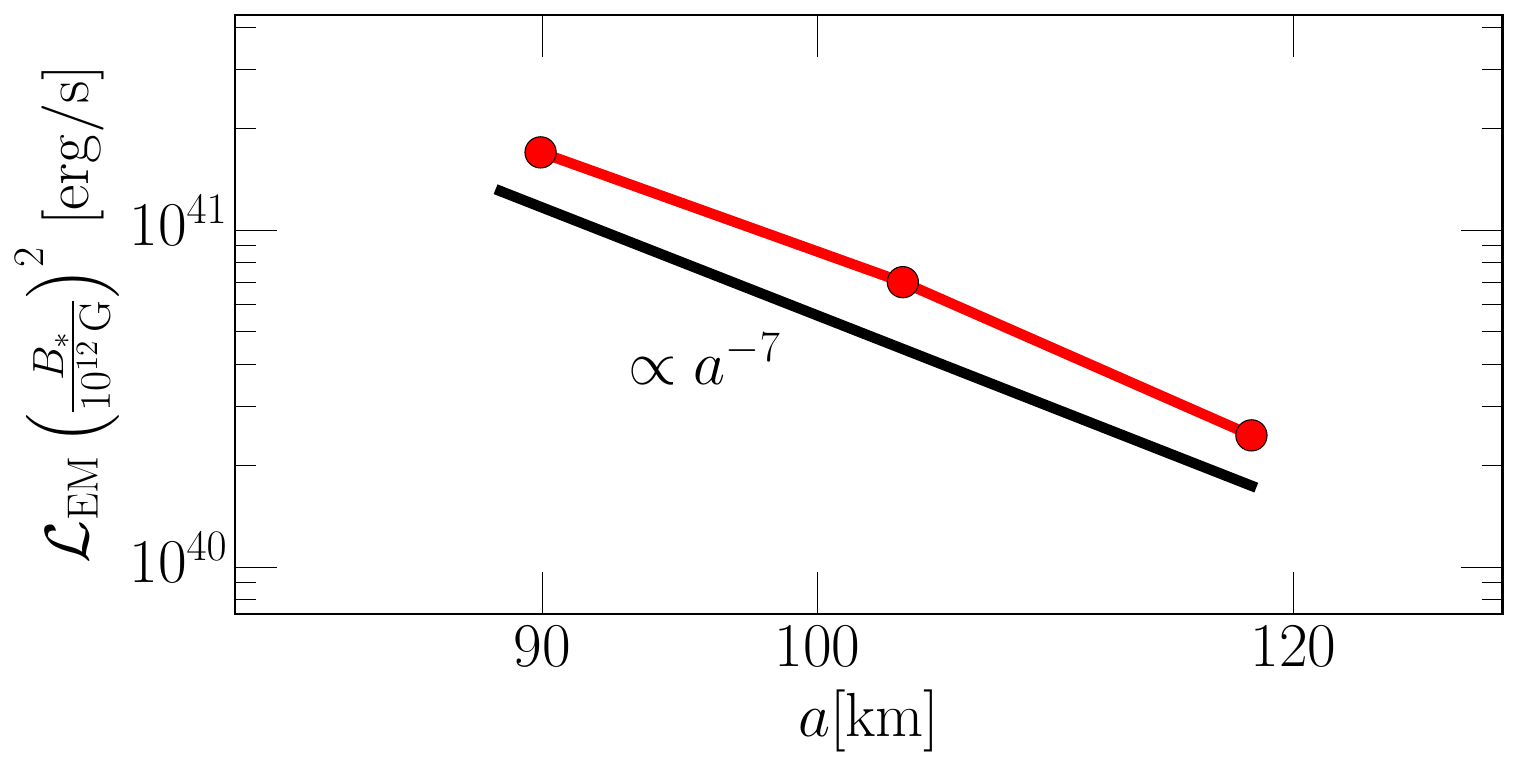}\\
    \caption{Bulk electromagnetic luminosity of the flares, $\mathcal{L}_{\rm EM}$. {\it (Top)} Scaling with black hole spin, $\chi_{\rm BH}$, relative to the luminosity, $\mathcal{L}_{\rm EM}^0$, of the aligned system.{\it (Bottom)} Dependence on orbital separation, $a$, with absolute magnitudes scaled to a surface field strength $B_\ast = 10^{12}\, \rm G$.}
    \label{fig:L_scaling}
\end{figure}

Having quantified the energetics of individual flares and their dependency
on the inclination angle $\theta_{\rm NS}$, we conclude this discussion by
clarifying the role of the other system parameters, notably (dimensionless)
BH spin, $\chi_{\rm BH}$, and orbital separation $a$.\footnote{NS spin would only affect the flaring periodicity but not the overall phenomenology or energetics estimates provided here \citep{Cherkis:2021vto}. This may be different in alternative scenarios not considered here \citep{Zhang:2019dpy,Dai:2019pgx}. 
Since NS are generally expected to be non-spinning at the time of merger
\citep{Bildsten:1992my}, we neglect NS spins in this work. The numerical infrastructure can, however, perfectly handle such a scenario (see, e.g., \citet{Most:2020exl,Tootle:2021umi,Papenfort:2022ywx}).}
\\ We first consider the impact of BH spin, $\chi_{\rm BH}$, as shown in
the upper panel of Fig.  \ref{fig:L_scaling}. We find that the flare
luminosities do not seem to depend on the BH spin. 
This is consistent with the flaring dynamics being governed by the orbital motion, not the extent of the ergosphere, which does depend on spin. It also matches earlier work on
constant emission from BH-NS systems \citep{Carrasco:2021jja}. In more
aligned cases, it might be possible that ergospheric motions for high spin
BHs could be able {to continuously twist connected flux tubes}, akin to what is expected in
models of magnetic loops connecting BHs and thin accretion flows \citep{Galeev:1979td,Uzdensky:2008ce,Yuan:2019mdb}. 
However, even when simulating models with BH spins of $\chi_{\rm BH} = 0.85$ with partially aligned or fully aligned magnetic moment we do not find signs of flaring. On the other hand,
our resistive force-free electrodynamics might be too diffusive compared to a fully
kinetic description of collisionless relativistic pair plasma \citep{Parfrey:2018dnc}.

We next focus on the dependency on the orbital separation, $a$. According to the estimate
presented for the moving dipole, Eq. \eqref{eqn:orbiting_dipole}, we expect the
luminosity to scale as $\mathcal{L}_{\rm EM} \propto a^{-7}$.
Comparing this with our simulations, we indeed find good agreement with
this scaling, in line with previous studies of continuous emission \citep{Carrasco:2021jja}.  
This strengthens the observation made above that
the flaring is mainly associated with the motion of the NS, rather than a
relative twist interaction as seen in the NS-NS case \citep{Most:2020ami}.
We caution that this scaling may be modified as the binary approaches merger,
i.e., on the last orbit. Full numerical relativity simulations of the
merger will be needed to clarify this point \citep{East:2021spd}.

Finally, we can also comment on the impact of mass ratio. While lighter BHs are astrophysically less likely due to the presumed existence of a lower mass gap (e.g., \citealt{Farr:2010tu,Ozel:2012ax}, but see also \cite{LIGOScientific:2020zkf} for a counterexample), it may be interesting to ask what the impact of more massive BHs in NS-BH system would be for the dynamics outlined here. To do so, we point out that the magnetospheric activity and its associated luminosity, $\mathcal{L}_{\rm EM}$, is governed by the orbital separation $a$, since $a$ determines the location relative to the NS's dipole field and therefore the strength of the field lines the BH is interacting with. To lowest order, the orbital separation will change over time, $t$, according to \citep{Peters:1964zz},
\begin{align}
    a \left(t\right) = a_0 \left(1- \frac{t}{t_{\rm merger}}\right)^{1/4}\,,
\end{align}
where $a_0$ is the initial separation with corresponding time to merger,
\begin{align}
t_{\rm merger} = \frac{5}{256} \frac{a_0^4}{m_{\rm NS}^3 Q \left(1+Q\right)}\,,
\end{align}
with, $Q=m_{\rm BH} / m_{\rm NS} >1$, being the mass ratio. This means that the time to merger, $t_{\rm merger} \sim \left(Q^2 +Q\right)^{-1}$, is shorter for more massive systems. Since the inspiral will accelerate towards the merger, the system will therefore spent fewer time in very close orbit (relative to the radial decay of the NS's magnetic field), and will -- in principle -- have EM outflows at lower luminosities. We quantify this explicitly in the next section. Since our simulations have BHs with masses close to the lower mass gap, our results can naturally be seen as upper bounds in terms of mass ratio dependence.

\subsection{Observational implications}
\label{sec:budget}

In the following, we want to provide a few simple estimates aimed at
assessing the potential observable impact of flares emitted from the
inspiralling NS-BH system.  We begin by estimating the number of
potentially observable flares, $N_{\rm flares}$, following our discussion in
\citet{Most:2022ojl}. Since the binary is emitting two flares per orbit, the
number of flares starting from a given separation, or equivalently from a
given orbital frequency, will be the same as the number of gravitational
wave cycles.  Using Kepler's law and the separation scaling consistent with the simulations (see Fig. \ref{fig:L_scaling}), we can estimate the frequency, $f_{\rm min}$, for
which the luminosity reaches a threshold value, $\mathcal{L}_{\rm min}$,
necessary for obtaining an observable signal.
Overall, we find
\begin{align}
    f_{\rm min} = \frac{1}{2 \pi} \frac{\sqrt{M}}{a_0^{3/2}}
    \left(\frac{\mathcal{L}_{\rm min}}{\mathcal{L}_0}\right)^{3/14}\,.
\end{align}
For an inspiralling compact binary, we can to lowest order estimate the number of gravitational wave cycles, $N_{\rm GW}$ \citep{maggiore2008gravitational}, 
\begin{align}
  N_{\rm flares} &= N_{\rm GW} \simeq \frac{1}{32 \pi^{8/3}} \left(\mathcal{M} f_{\rm min}\right)^{-5/3}\,,\\
    &\simeq 2\ \left(\frac{B}{10^{12}\, \rm G}\right)^{5/7} \left(\frac{10^{42}\ \rm erg/s}{\mathcal{L}_{\rm min}}\right)^{5/14} \left(\frac{\sigma}{10^{-3}}\right)^{5/14}\,, \label{eqn:nflares}
\end{align}
where the latter is obtained for a GW200115-like system. We have further
introduced a parameter, $\sigma=10^{-3}$, estimating the efficiency of
converting the outgoing luminosity into observable radiation (see \citet{Most:2022ojl} for a discussion). We point out that
this estimate is about ten-times lower than in the binary neutron star case
\citep{Most:2022ojl}, but can reach the same order, if NS field strengths
$B_\ast \simeq 10^{13}\, \rm G$ are present. This finding is overall
consistent with the fact, that the last orbits prior to merger will happen
at almost twice the orbital separation compared to the NS-NS case. The
flaring process of outer field lines then in turn will only involve outer,
lower field strengths regions of the closed NS magnetosphere. In addition, NS-NS merger flaring features a shallower separation dependence, $\mathcal{L}^{\rm NS-NS}_{\rm EM} \sim a^{-3/5}$ \citep{Most:2020ami}. 
We can also explicitly quantify the dependence on the mass ratio, $Q$, which we find to be $N_{\rm flares} \sim Q^{-1} \left(1 +Q\right)^{-1/2}$. This implies that flares may not leave observable signatures for larger mass ratios with $N_{\rm flares} \lesssim 1$ for $Q\gtrsim 6$. \\

\noindent {\it How do flares convert their energy into transients?} \\
Having clarified the number of flares expected from the late inspiral of a
realistic BH-NS binary, we now discuss the potential emission channels
open to converting the energy of the flares into observable transients.

First, we focus on the production through forced reconnection in the
orbital current sheet.
Reconnection is enhanced due to forcing in the outer parts of the
field lines, leading to the copious ejection of plasmoids. The merger of
these plasmoids can convert a sizeable fraction of the available energy into low frequency electromagntic waves \citep{Philippov:2019qud}. Dissipation of magnetic energy due to reconnection in the current sheet leads to X-ray emission  \citep{Beloborodov:2020ylo}.
Although resistive
dissipation can only be approximately captured in our simulations, we
conservatively measure an upper bound of $\mathcal{L}_{\rm diss} \lesssim
0.3 \mathcal{L}_{\rm EM}$, consistent with our previous work on NS-NS
systems \citep{Most:2020ami,Most:2022ojl}. \\
Focusing on the flares, two main scenarios of the low frequency coherent emission are possible. 
Either the flares propagate out to large distances, where they will
interact with the binaries pair wind. Such a scenario has been shown to
be capable of producing Fast Radio Bursts in the case of magnetars
\citep{Beloborodov:2017juh,Metzger:2019una}, albeit we here are faced with the prospect of having far less
energetical flares. \\
In the second scenario, the flares can depending on the inclination angle, $\theta_{\rm NS}$, of the NS's magnetic moment be
emitted along the equatorial plane (see Fig. \ref{fig:2D_poynting}), where they
will eventually collide with the orbital current sheet. The compression of
the sheet results in the formation of plasmoids and the launching of
small-scale fast waves with wavelength potentially in the radio band
\citep{Lyubarsky:2020qbp}. Such a scenario has been proposed for FRB
emission in magnetars and has recently been investigated numerically using
kinetic simulations \citep{Mahlmann:2022nnz}. 
We caution that interacting compact binaries may produce FRB-like transients
also by other means (for a comprehensive recent review see, e.g., \citet{Zhang:2022uzl}).
Since some of these will depend on kinetic or magnetohydrodynamic (MHD) physics (i.e., finite magnetization and shock formation) not included in our force-free simulations, 
we do not consider them here. 
In an earlier work, we have numerically investigated the flare
collision dynamics for NS-NS systems, showing the feasibility of this
mechanism to produce Fast Radio Burst-like transients \citep{Most:2022ayk}.\\
In the following, we will apply this model to the BH-NS parameters
extracted from our simulations as well. 
We begin by computing the effective Lorentz factor, $\Gamma$, of the plasma in the
flare. Following \citet{Lyubarsky:2020qbp}, we can estimate $\Gamma = \sqrt{B_{\rm
flare}/B_{\rm wind}} \approx 1.6$, where we have used that in our
simulation the flaring field strength $B_{\rm flare} = 5 \times 10^8\,
B_{12}^\ast\, \rm G$ ist ten-times larger than the wind strength $B_{\rm
wind}$, where $B_{12}^\ast= B_\ast / 10^{12}\, \rm G$ is the surface magnetic field strength. As the flare
collides with the current sheet, it triggers enhanced plasmoid formation and production of low frequency fast magnetosonic waves. In highly magnetized plasma these waves can propagate even at frequencies below the plasma frequency. Morever, since they propagate on top of the wind, they avoid damping expected for GHz waves in the inner magnetosphere \citep{Beloborodov:2023lxl}.
The frequency of the resulting emission
due to plasmoid collision is then inversely proportional to the plasmoid
size \citep{Lyubarsky:2020qbp}. Since our simulations do not correctly capture collisionless
reconnection physics, we translate results of small-scale kinetic
simulations \citep{Mahlmann:2022nnz} to compact binary magnetospheres
\citep{Most:2022ayk}.
For this scenario, it can then be shown that the characteristic emission
frequency \citep{Lyubarsky:2020qbp},
\begin{align}
  \nu_{\rm FRB} =  \frac{1}{2 \pi \xi \zeta}
  \sqrt{\frac{2 r_e}{3 \beta_{\rm rec} c \Gamma} \omega_B^3} \simeq 9 \,
  \left(B^\ast_{12}\right)^{3/2}\,\rm GHz\,.
  \label{eqn:nu_flare}
\end{align}
Here we have introduced the electron gyrofrequency in the flare via
$\omega_B = e B_{\rm flare} / m_e c$, where $e$, $m_e$ and $r_e$ are the
electron charge, mass and classical radius, respectively. We have also used
$\xi \zeta \approx 100$
\citep{Mahlmann:2022nnz}, as well as a reconnection rate $\beta_{\rm rec}\simeq 0.1$
\citep{Sironi:2014jfa,Guo:2015cua}.
These estimates imply that for surface field strength $B_\ast \simeq 10^{12}\, \rm G$,
Fast-Radio-Burst-like\footnote{We recall that due to the millisecond
orbital time scale, the burst time will be of a similar order.} transients
can indeed be launched, albeit only for about two flaring events, likely happening in the last orbit, see Eq. \eqref{eqn:nflares}. 
Conversely, if the field
strength was increased to achieve multiple flaring episodes, these would
quickly go outside of the radio band. We caution that the exact frequency value
will strongly depend on the radial distance at which the flare hits the
orbital current sheet (see \citet{Most:2022ayk} for a discussion).

\section{Conclusions}\label{sec:conclusions}

Apart from gravitational wave emission, compact binary inspirals could
 be accompanied by electromagnetic precursors sourced prior to merger
 \citep{Hansen:2000am,Lai:2012qe,Piro:2012rq,Lyutikov:2018nti}. These could be produced as a result of dynamical interactions 
 in the common magnetosphere
 \citep{Lai:2012qe,Piro:2012rq,Palenzuela:2013hu,Palenzuela:2013kra,Ponce:2014hha,Most:2020ami,Most:2022ojl}, tidal deformation of the NSs crust
 \citep{Tsang:2011ad}, or as a result of partially expelling the magnetosphere after the merger (balding transient)
 \citep{Nathanail:2020fkp,East:2021spd}. All these scenarios have in common that they will feature some
 dependence on the magnetic field topology, and the spin present in the
 initial system. Since they are either not well constrainable from the
 gravitational wave signal \cite{}, or in the case of the magnetic field
 topology might get destroyed during merger \citep{Aguilera-Miret:2020dhz}, 
 such precursors might
 offer one of the few glimpses into the system. Even more so, sky localization
 of gravitational wave events currently suffers from large uncertainties
 impeding swift follow-up observations. If radio transients were sourced
 by the merger, this would potentially enable fast localization by means of
 all-sky radio observations \citep{Sachdev:2020lfd,Wang:2020sda,Yu:2021vvm}.\\
 In this work, we have investigated such a scenario for BH-NS binaries, in
 particular realistic configurations for which likely no remnant baryon
 mass, which could power additional transient, will be present.
 By means of general-relativistic force-free electrodynamic simulations we
 have demonstrated that relative orbital motions will naturally form closed
 magnetic flux tubes, the twisting of which causes them to erupt akin to
 solar coronal mass ejections \citep{chen2011coronal}. %
 We find that the emission
 of flares correlates with the part of the closed magnetosphere closest to the orbital light cylinder. These field lines may be sheared more easily, leading to an easier flaring. More specifically, we find that
 magnetic field inclinations of $45^\circ$ and larger will lead to strong
 outbursts. For aligned systems, we instead find that the black hole
 continuously dissipated energy establishing a steady state twist. This effect seems to get enhanced by the orbital motion, rather than by BH spin, which instead controls the size of the ergosphere. We caution that the precise picture
 might change when collisionless reconnection physics is properly
 accounted for, either by means of kinetic \citep{Philippov:2014mqa,Kalapotharakos:2017bpx}, 
 two-fluid simulations \citep{wang2018electron,dong2019global} or even in ideal MHD, necessitating novel formulations in the relativistic context \citep{Most:2021uck}. \\
 On the quantitative side, we find that flares manage to significantly enhance the
 electromagnetic luminosity compared to the orbiting dipole estimate
 \citep{Ioka:2000yb,Hansen:2000am}. This is in line with previous work on BH-NS \citep{Carrasco:2021jja} and NS-NS
 magnetospheres \cite{Most:2020ami,Most:2022ojl}. Concretely, we find that magnetic field strengths
 of $B_\ast = 10^{12}\, \rm G$ at the stellar surface, correspond to
 luminosities around $\mathcal{L}_{\rm EM} \simeq 10^{41}\, \rm erg /s$. 
 This scales strongly with the separation (but not BH spin), consistent with previous
 analytical and numerical calculations for continuous electromagnetic
 emission from BH-NS mergers \citep{Carrasco:2021jja}. Overall, we estimate that this leads
 to at most two potentially detectable flares when using the parameters
 listed above. However, their emission, if driven by current sheet -- flare
 interactions \citep{Lyubarsky:2020qbp,Mahlmann:2021yws,Most:2022ayk}, will likely correspond to that of
 fast radio burst-like transients, peaking in the $\nu = 9\, \rm GHz$
 range, making them interesting targets for all-sky radio observatories
 \citep{Callister:2019biq,James:2019xca}. More precise studies, which also account for systems with orbital precession
 will be needed to clarify the uncertainties in the present estimates,
 which largely depend on the radial location of the interaction of flares with the
 current sheet.
 While this work has largely considered flares as the main source of EM transients, larger Poynting flux outflows can also be produced at merger \citep{,DOrazio:2013ngp,Mingarelli:2015bpo,DOrazio:2015jcb,Nathanail:2020fkp,East:2021spd}. However, it remains uncertain how exactly these outflows would dissipate and, hence, what exact observable signatures they would correspond to. We plan to address such configurations in a future work.
 
 The main challenge compared to NS-NS systems fundamentally
 lies in the expected mass ratio of these sytems \citep{Fragione:2021cvv,Biscoveanu:2022iue}. Since the binary
 will orbit at larger separations compared to an equal mass system, the
 BH will fundamentally only interact with the outermost field lines of the
 NS, which flare more easily. Whether or not interactions with
 field lines closer to the star could lead to flaring remains unclear, due
 to the dissipative nature of the resistive force-free approach \citep{Mahlmann:2020yxn}. Future work
 will be needed to clarify these points.\\

\section*{Acknowledgments}
The authors are grateful for discussions with Benoit Cerutti, Sean McWilliams and Navin Sridhar.
ERM acknowledges by the National Science Foundation under Grant No. AST-2307394.
AP acknowledges support by the National Science Foundation under grant No. AST-1909458.
This work was initiated during visits at the Aspen Center for
Physics, which is supported by National Science Foundation grant
PHY-1607611, and at the Institute for Computational and Experimental Research in Mathematics in Providence, RI, 
which is supported by the National Science Foundation under Grant No. DMS-1929284.
This research was facilitated by the Multimessenger Plasma Physics Center (MPPC), NSF grant PHY-2206610. The simulations were performed on the NSF Frontera supercomputer under grant AST21006. 
ERM acknowledges the use of Delta at the National Center for Supercomputing Applications (NCSA) through allocation PHY210074 from the Advanced Cyberinfrastructure Coordination Ecosystem: Services \& Support (ACCESS) program, which is supported by National Science Foundation grants \#2138259, \#2138286, \#2138307, \#2137603, and \#2138296. 
Support also comes from the Resnick High Performance Computing Center, a facility supported by Resnick Sustainability Institute at the California Institute of Technology.

\software{AMReX \citep{amrex},
	  FUKA \citep{Papenfort:2021hod},
	  Kadath \citep{Grandclement:2009ju},
	  matplotlib \citep{Hunter:2007},
	  numpy \citep{harris2020array},
	  scipy \citep{2020SciPy-NMeth}
}

\bibliography{inspire,non_inspire}

\end{document}